\documentclass[prl,twocolumn,showpacs,preprintnumbers,superscriptaddress,amsmath,amssymb]{revtex4}

\usepackage{epsf,color,graphicx}
\usepackage[dvips]{epsfig}

\begin{document}

\title{Ultra-low threshold polariton lasing in photonic crystal cavities}

\author{Stefano Azzini}
\author{Dario Gerace}
\author{Matteo Galli}
\affiliation{Dipartimento di Fisica ``A. Volta," and UdR CNISM,  via Bassi 6, 27100 Pavia, Italy}
\author{Isabelle Sagnes}
\author{R\'{e}my Braive}
\author{Aristide Lema\^{i}tre}
\author{Jacqueline Bloch}
\affiliation{CNRS-Laboratoire de Photonique et Nanostructures, Route
de Nozay, 91460 Marcoussis, France}
\author{D. Bajoni}\email[]{daniele.bajoni@unipv.it}
\affiliation{Dipartimento di Elettronica and UdR CNISM, via Ferrata 1, 27100 Pavia, Italy}

\date{\today}

\begin{abstract}
The authors show clear experimental evidence of lasing of exciton polaritons confined 
in L3 photonic crystal cavities. 
The samples are based on an InP membrane in air containing five InAsP quantum wells. 
Polariton lasing is observed with thresholds as low as 120 nW, below the Mott transition, 
while conventional photon lasing is observed for a pumping power one to three orders of 
magnitude higher.
\end{abstract}


\maketitle


Polariton lasing originates from the spontaneous formation of a coherent population of exciton-polaritons 
out of incoherent excitation~\cite{Imamoglu}. 
Exciton-polaritons are the dressed states arising from the strong coupling of a photonic mode in 
a semiconductor microcavity with excitons confined in an embedded quantum well (QW)~\cite{SkolnickSST}. 
Polariton lasers act as coherent light sources very similar to conventional lasers,
the main difference being that polariton lasing occurs below the pumping 
rates necessary for population inversion: the formation mechanism of the coherent polariton state 
is stimulated relaxation of polaritons~\cite{Malpuech}, as opposed to stimulated emission of 
photons. As a result, the threshold for polariton lasing has been predicted~\cite{Porras} and 
observed to be several orders of magnitude below the conventional photon lasing threshold 
in the same samples~\cite{Deveaud,Bajonilasing}, and recently reported up to room 
temperature in GaN based samples~\cite{Grandjean}.
However, the threshold powers for polariton lasing reported up to date are larger than (or comparable 
to) the lowest threshold reported for conventional lasers obtained with the same materials~\cite{VCSELs}. 
This is mainly due to the inability to confine polaritons in volumes comparable 
to their optical wavelength: polariton ``boxes'' such as micropillars and cavity 
corrugations have been reported with confinement volumes on the order of 
tens of $\mu$m$^3$\cite{boxes}.

To date, photonic nanocavities realized by point defects in photonic crystal (PC) slabs \cite{Joannopoulos}can be fabricated by top-down lithographic techniques \cite{Lith}, yielding unprecedented figures of merit in terms of quality factor (Q) over effective confinement volume (V$_{\mathrm{eff}}$) \cite{highQ}. The typical L3 cavity design \cite{Noda}, with three missing holes along the $\Gamma K$ direction in a triangular lattice, supports diffraction limited cavity modes $V_{\mathrm{eff}}\simeq (\lambda /n)^3$, allowing the demonstration of basic cavity QED effects \cite{Englund,Badolato} and ultra-low threshold lasing \cite{Hennessy}.

Despite an ongoing research effort to observe the strong coupling regime in PC structures with embedded QWs \cite{Gerace}, the structures realized so far rely on the periodic modulation of the evanescent tail in the photonic mode \cite{BajoniPCPol,Zanotto}. This is mainly due to the introduction of fast recombination channels for the excitons when patterning GaAs based QWs, which hinders exciton coherence to point of preventing strong coupling \cite{BajoniPCPol}. In this work we choose InP-based materials for their negligible nonradiative recombination issues, even after patterning, and we report experimental evidence of polariton lasing in L3 photonic crystal cavities.  

\begin{figure}[t!]
\includegraphics[width= 0.95\columnwidth]{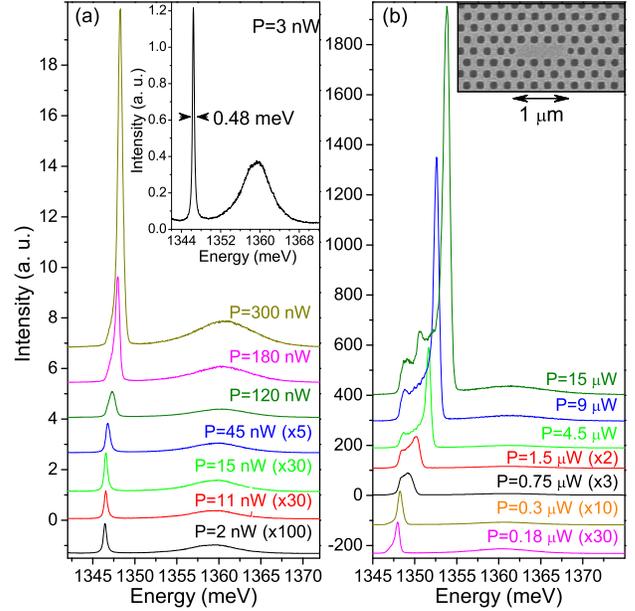}
\caption{(color online) (a) PL spectra measured on a sample with $a=235$ nm for increasing pump
power $P$; in the inset a detail of the spectrum for $P= 3$ nW is shown. 
(b) PL spectra measured on the same sample at higher pumping; in the inset a SEM 
image of the cavity is reported.}\label{Fig1}
\end{figure}

The cavities have been fabricated on an InP suspended membrane. A 230 nm-thick InP 
guiding layer was grown by molecular beam epitaxy on top of a 1.5 $\mu$m thick InGaAs sacrificial layer, 
on InP substrate. The InP guiding layer contains five $8$ nm-thick shallow InAsP QWs separated by 12 nm InP 
barriers at its center.  The L3 cavities have 
been obtained via standard electron beam lithography followed by inductively coupled plasma dry etching 
of the InP top layer. After the etching, the 1.5 $\mu$m sacrificial layer was selectively removed by wet etching 
to produce air suspended membranes. The structural parameters of the PCs  were chosen to 
have the resonance condition between the fundamental L3 cavity mode 
and the QW s-wave exciton. By lithographic tuning, the lattice constant $a$ was scanned between 
230 and 250 nm every 5 nm, while the ratio between the hole's radius and the lattice constant is kept fixed as $r/a=0.32$. 
The holes at specific positions around the cavity were slightly varied in size to maximize the 
out-of-plane emission from the cavity mode~\cite{Tran}. 
A SEM picture of the cavity region is shown in the inset of Fig.~1(b). Photoluminescence (PL) experiments were 
carried out exciting the samples with a pulsed laser pump (10 ps pulse width) at $\lambda=$750 nm 
focussed on a 500 nm spot through a high numerical aperture microscope objective, and the PL signal was 
selectively collected from the cavity using a confocal set-up through the same objective. The sample was kept 
at 10 K in a cold finger cryostat.

\begin{figure}[t!]
\includegraphics[width= 0.96\columnwidth]{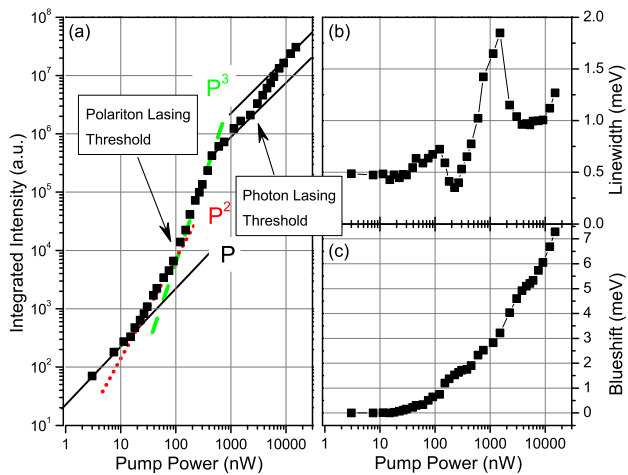}
\caption{(color online) (a) Integrated intensity of the mode from PL spectra shown in Fig. 1 as a function of 
the pump power. The continuous (black) line, the dotted (red) line and the dashed (green) line are guides 
to the eye proportional to, respectively, the pump power $P$, $P^2$ and $P^3$. (b) Linewidth and (c) blueshift 
the mode from PL spectra shown in Fig. 1.}\label{Fig2}
\end{figure}

A typical PL spectrum from the cavities is shown in the inset of Fig.~1(a). Two main features can be highlighted: 
a broad resonance at 1360 meV, visible also outside the patterned area, due to the emission from bare QW excitons, and a sharp resonance
on the low energy side of the exciton transition. 
The quality factor of these polariton resonances was, in all samples, between 3000 and 6000, corresponding to a 
lifetime of the order of a ps.
Spectra taken from a sample with $a=235$ nm are reported in Figs.~1(a) and (b) for increasing pump power, $P$. 
The PL emission shows a clearly nonlinear behavior: when the excitation power is increased above $P\sim 100$ 
nW an evident blueshift and a super-linear increase of PL from the polariton line can be
observed in Fig.~1(a). Another similarly nonlinear threshold, accompanied by an even larger blueshift 
is observed for $P>2$ $\mu$W in Fig~1(b). Between these two thresholds, the line significantly broadens. 

We summarize in Fig.~2 the behaviors of the integrated peak intensity, its linewidth and blueshift, 
respectively, as a function of pump power. Both the first and the second 
thresholds are accompanied by a spectral narrowing of the emission, implying the increase of temporal 
coherence. Within the first threshold the emitted peak shifts by about 1 meV, while a total shift 
of more than 5 meV is observed before the onset of the second threshold.
The presence of both thresholds is an unambiguous (although indirect) evidence that the sample is in strong 
coupling at  low pumping powers, and that we are indeed in presence of both polariton lasing 
(with threshold around $P\sim 100$ nW, corresponding to a power density of $\sim 50$ W/cm$^2$), and conventional photon lasing (with threshold around  $P\sim 2$ $\mu$W, $\sim 1$ kW/cm$^2$). 

Unfortunately, it was not possible to observe anticrossing between exciton and bare cavity mode. 
In fact, temperature cannot be used as a tuning parameter, as the InAsP QW exciton shifts by less than 1 nm 
between 4 K and 70 K, while the cavity mode shifts by less than 2 nm using thin film coating in the cryostat. 
Moreover, lithographic tuning is too coarse and the points too few to be used for a reliable anticrossing plot. 
However, we stress that the presence of two thresholds, separated by the Mott transition, is a sufficient 
proof that the sample is in strong coupling for pumping powers below 1 $\mu$W.
This is also confirmed by the blueshift, which continues well above the first threshold. This is a clear
indication that the sample is entering the weak coupling regime, and the emission resonance is shifting 
from the lower polariton to the bare cavity mode~\cite{Bajonilasing}.
Notice that just below polariton lasing (between 30 nW and 100 nW) there is a quadratic increase in the 
emission intensity: such a dependence is the fingerprint  that the dominating relaxation mechanism giving
rise to polariton lasing is polariton-polariton scattering, as predicted~\cite{Porras}.
Notice also that the lasing threshold in these samples is reduced by three orders of magnitude with
respect to the existing literature~\cite{Bajonilasing,Grandjean,Deveaud}, and is comparable to the lowest 
thresholds reported for quantum dots lasers~\cite{Hennessy} so far. The threshold for photon lasing, on the 
contrary, occurs at powers consistent to those reported for other InP-based PC cavity lasers ~\cite{Postigo}.

We have observed polariton lasing in samples with a different lattice constant, and thus different exciton/cavity 
detuning, $\Delta=E_{\mathrm{cav}} - E_{\mathrm{exc}} $. 
The threshold power increases with increasing $\Delta$, hence with the photonic component of the polariton, 
as expected. Polariton lasing relies on polariton-polariton scattering, so it is strongly dependent on the 
excitonic fraction. In Fig.~3 we report PL spectra collected for increasing pump power on a sample with lattice
constant $a=250$ nm. In this case, $\Delta$ is too large and the exciton fraction is not enough to obtain 
polariton lasing: as it is shown in Fig.~3(c) the emitted intensity increases linearly, while the line broadens 
and blueshifts due to the progressive loss of strong coupling. 
The crossover to weak coupling is observed around $P\sim 1$ $\mu$W as in all other samples. 
When the sample is in weak coupling the blueshift stops, and for $P>10$ $\mu$W conventional
photon lasing sets in with a super-linear increase of the emitted intensity. The fact that the first threshold is not observed far from the exciton resonance proves it is due to excitonic gain, and not to conventional gain due to band filling.

At such large negative detunings, changes in refractive
index with pumping related to the exciton resonance are negligible
 \cite{PhotonLasing}. However effects due to the injected electron-hole pairs have to be taken
into account following Ref. \onlinecite{Fushman} and using InP parameters \cite{InP}. We
obtain that the bare cavity mode is blueshifted by $\sim$2.5 meV, which
gives a bare cavity resonance at $E_{\mathrm{cav}}=1329.7$ meV. Knowing that the QW exciton energy is
$E_{\mathrm{exc}}=1360$ meV, and measuring the lower polariton energy below threshold as 
$E_{\mathrm{LP}}=1328.3$ meV, we can estimate  the value of the Rabi splitting from a simple
two oscillators model as $\hbar \Omega=\sqrt{\left(2 E_{\mathrm{LP}}-E_{\mathrm{cav}}-E_{\mathrm{exc}} \right)^2-\Delta^2}\simeq 
13.5$ meV~\cite{SkolnickSST} (we estimate an uncertainty of $\sim$ 1 meV on this value). This value of $\hbar \Omega$ is consistent with what expected for five GaAs-based QWs in a comparable system \cite{Gerace}. The correspondent detuning is $\Delta\simeq -10$ meV for the sample of Figs.~1 and 2 
(i.e. $a=235$ nm), and $\Delta\simeq -31$ meV for the sample of Fig. 3 ($a=250$ nm).

In conclusion the reduction of the modal volume with respect to previously studied solutions for polariton 
confinement leads to a reduction of more than three orders of magnitude in polariton lasing threshold. 
The ability to confine polaritons in volumes comparable to cube of their wavelength should also enable
to observe effects related to the enhancement of their repulsion, such as polariton self-phase 
modulation~\cite{Whittaker} and ultimately polariton blockade \cite{Ciuti}.

This work was supported by CNISM funding through the INNESCO project ÒPcPolÓ, by MIUR funding 
through the FIRB ``Futuro in Ricerca" project RBFR08XMVY and from the foundation Alma Mater Ticinensis. We acknowledge L. C. Andreani and M. Patrini for fruitful discussions and L. Ferrera for a first characterization of the sample.

\begin{figure}[t!]
\includegraphics[width= 0.95\columnwidth]{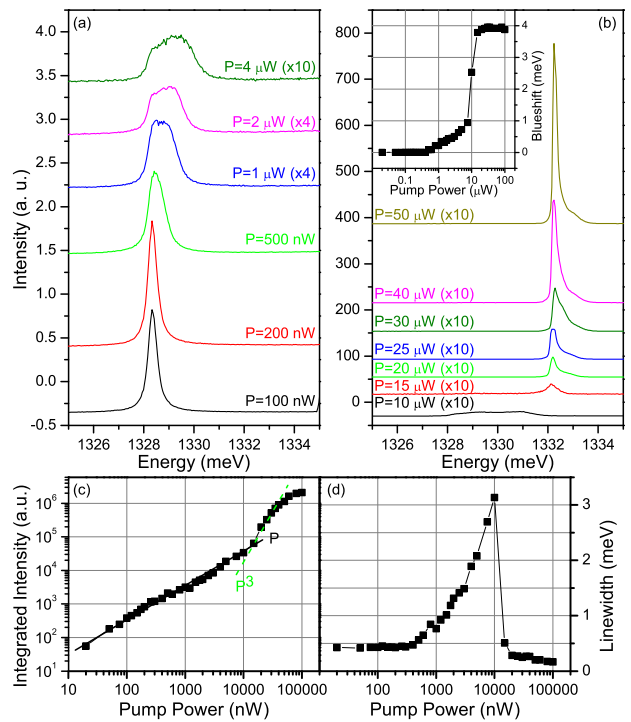}
\caption{(a) and (b) PL spectra measured on a sample with $a=250$ nm for increasing P; inset: blueshift 
extracted from the data. (c) Integrated intensity as a function of the pump power. The continuous (black)
 line and the dashed (green) line are guides to the eye proportional to, respectively, the pump power P 
 and P$^3$. (d) Linewidth from the data of panels (a) and (b)}\label{Fig3}
\end{figure}

\end{document}